\documentclass[prb,twocolumn,showpacs,preprintnumbers,amsmath,amssymb,superscriptaddress]{revtex4}
\usepackage{dcolumn}
\usepackage{bm,graphicx}

\begin{document}

\title{Magneto-optics of bilayer graphene inclusions in rotational-stacked \\ multilayer epitaxial graphene}
\author{M. Orlita}\email{milan.orlita@lncmi.cnrs.fr}
\altaffiliation{also at Institute of Physics, v.v.i., ASCR, Prague, Czech Republic}
\affiliation{Laboratoire
National des Champs Magn\'etiques Intenses, CNRS-UJF-UPS-INSA, 25, avenue des
Martyrs, 38042 Grenoble, France}
\affiliation{Institute of Physics, Faculty of Mathematics and Physics, Charles University, Ke Karlovu 5,
121~16 Praha 2, Czech Republic}
\author{C. Faugeras}
\affiliation{Laboratoire National des Champs Magn\'etiques Intenses,
CNRS-UJF-UPS-INSA, 25, avenue des Martyrs, 38042 Grenoble, France}
\author{J. Borysiuk}
\affiliation{Institute of Physics, Polish Academy of Sciences, 02-668 Warsaw, Al.
Lotnikow 32/46, Poland}
\author{J. M. Baranowski}
\affiliation{Institute of Experimental Physics, University of Warsaw, Ho\.{z}a
69, PL 00-681 Warsaw, Poland}
\author{W.~Strupi\'{n}ski}
\affiliation{Institute of Electronic Materials Technology, PL 01-919 Warsaw,
Poland}
\author{M.~Sprinkle}
\affiliation{School of Physics, Georgia Institute of Technology, Atlanta,
Georgia 30332, USA}
\author{C.~Berger}
\affiliation{School of Physics, Georgia Institute of Technology, Atlanta,
Georgia 30332, USA} \affiliation{Institut N\'{e}el/CNRS-UJF BP 166, F-38042
Grenoble Cedex 9, France}
\author{W. A. de Heer}
\affiliation{School of Physics, Georgia Institute of Technology, Atlanta,
Georgia 30332, USA}
\author{D. M. Basko}
\affiliation{Laboratoire de Physique et Mod\'elisation des Milieux Condens\'es,
UJF and CNRS, F-38042 Grenoble, France}
\author{G. Martinez}
\affiliation{Laboratoire National des Champs Magn\'etiques Intenses,
CNRS-UJF-UPS-INSA, 25, avenue des Martyrs, 38042 Grenoble, France}
\author{M. Potemski}
\affiliation{Laboratoire National des Champs Magn\'etiques Intenses,
CNRS-UJF-UPS-INSA, 25, avenue des Martyrs, 38042 Grenoble, France}
\date{\today}

\begin{abstract}
Additional component in multi-layer epitaxial graphene grown on
the C-terminated surface of SiC, which exhibits the characteristic
electronic properties of a AB-stacked graphene bilayer, is
identified in magneto-optical response of this material. We show
that these inclusions represent a well-defined platform  for
accurate magneto-spectroscopy of unperturbed graphene bilayers.
\end{abstract}


\maketitle

\section{Introduction}

The absence of the energetic gap in the excitation spectrum of
graphene\cite{NovoselovNM07} is considered as a possible drawback
preventing the straightforward application of this emerging
material in electronics. This is despite numerous efforts, such as
those implying surface patterning\cite{BaiNatureNano10} and
substrate- or adsorbents-induced
interactions.\cite{ZhouNatureMater07,BostwickNaturePhys07,BalogNatureNano10}
The possibility to open and tune the band gap in the bilayer
graphene has recently been demonstrated by applying an electric
field perpendicular to the graphitic
planes\cite{KuzmenkoPRB09,MakPRL09,ZhangNature09} and this is a
key element to construct a transistor, the building block of
electronic circuits. The band gap engineering is ``typical'' of
the bilayer and is not reported in tri- and more-layer graphene
specimens where semi-metallic behavior
dominates.\cite{CraciunNatureNano09} From the viewpoint of
applications, the bilayer graphene thus becomes almost equally
appealing material as graphene itself.

Optical spectroscopy has played an important role in
investigations of the bilayer
graphene,\cite{ZhangPRB08,ZQLiPRL09,KuzmenkoPRB09} as, for
instance, it allows to directly visualize the electric-field
induced energy gap in this system.\cite{ZhangNature09,MakPRL09} On
the other hand, only relatively scarce information has been up to
now collected from \textit{magneto}-optical
measurements.\cite{HenriksenPRL08} This fact might be surprising
when noticing the potential of Landau level (LL) spectroscopy
which has been widely applied to other graphene-like systems.
Magneto-optical methods have, for example, convincingly
illustrated the unconventional LL spectrum in graphene, have
offered a reliable estimate of the Fermi velocity or invoked the
specific effects of many-body interactions between massless Dirac
fermions.\cite{SadowskiPRL06,JiangPRL07,DeaconPRB07,OrlitaPRL08II,NeugebauerPRL09,HenriksenPRL10,CrasseeCM10}

So far, the only magneto-optical experiments on the bilayer
graphene have been reported by Henriksen \textit{et
al.},\cite{HenriksenPRL08} who succeeded to probe a relatively
weak cyclotron-resonance signal of a small flake using the
gate-controlled differential technique. The optical response at a
fixed magnetic field was then studied as a function of the carrier
density. Such differential spectroscopy was efficient in case of
exfoliated graphene monolayers,\cite{JiangPRL07,HenriksenPRL10}
but it provides more complex results when applied to the bilayer
graphene. In this latter system, the change of the gate voltage
affects not only the carrier density but also modifies
significantly the band structure and data interpretation is by far
more elaborated.\cite{MuchaKruczynskiSSC09}


In this paper, we demonstrate that certain class of previously
reported AB-stacking
faults\cite{FaugerasAPL08,SprinklePRL09,SprinkleCM10,SiegelPRB10}
in otherwise rotationally-ordered multilayer epitaxial graphene
(MEG),\cite{BergerJPCB04,HassPRL08,SadowskiPRL06,MillerScience09}
show the characteristic features of well-defined graphene
bilayers. These inclusions, identified here in
magneto-transmission experiments, represent therefore a suitable
system for accurate magneto-spectroscopy studies of
unperturbed bilayer graphene.

\begin{figure*}
    \begin{minipage}{0.6\linewidth}
      \scalebox{.9}{\includegraphics{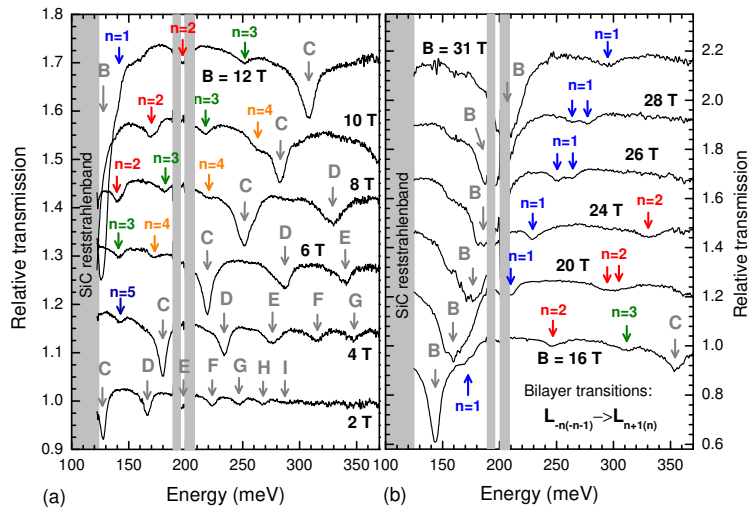}}
    \end{minipage}\hfill
    \begin{minipage}{0.32\linewidth}
      \caption{\label{SPKT} (color online) Transmission spectra of MEG with $\sim$100 layers recorded at the
selected magnetic fields below 12 and above 16~T in parts (a) and (b),
respectively, in both cases above the reststrahlenband of SiC. Whereas the
transitions marked B to I correspond to electrically isolated graphene
monolayers, transitions denoted $n=1$ to 5 match to inter-LL transitions in an
unperturbed graphene bilayer, following the coding
L$_{-n(-n-1)}\rightarrow$L$_{n+1(n)}$. For clarity, successive spectra in parts (a) and (b) are shifted
vertically by 0.14 and 0.23, respectively.}
    \end{minipage}
\end{figure*}

\section{Sample preparation and experimental details}

The growth of MEG samples studied here was performed with a
commercially available horizontal chemical vapor deposition
hot-wall reactor (Epigress V508), inductively heated by a RF
generator. Epitaxial MEG films were grown on semi-insulating
4H-SiC(000\={1}) on-axis C-terminated substrates at 1600$^\circ$C
in Ar atmosphere. The growth rate was controlled by the Ar
pressure ($\sim$100~mbar) which was found to directly influence
the evaporation rate of Si atoms.

To measure the infrared transmittance of our samples, we used the
radiation of a globar, which was analyzed by a Fourier transform
spectrometer and delivered to the sample via light-pipe optics.
The transmitted light was detected by a composite bolometer kept
at $T=2$~K and placed directly below the sample. Measurements were
carried out in superconducting ($B= 0 - 13$~T) and resistive ($B=
13 - 32$~T) solenoids with spectral resolution of 0.5 and 1~meV in
the range of magnetic field below and above $B= 13$~T,
respectively. All presented spectra were normalized by the sample
transmission at $B=0$.

The samples were characterized in micro-Raman scattering
experiments which, similarly to previous
studies,\cite{FaugerasAPL08} revealed, depending on location, single-component 2D band
features, characteristic of graphene simple electronic bands and
of decoupled graphitic planes in multilayer epitaxial graphene
grown on the C-face of a SiC substrate, or multi-component
2D band features characteristic of Bernal stacked graphite. In
this paper, we present transmission data obtained on one
particular specimen with a high number of graphitic layers
($\sim100$) and grown on a SiC substrate with a reduced thickness
of $\sim100$~$\mu$m. Due to this latter condition, the spectral
region of total opacity of the sample only covers the SiC
reststrahlenband ($\sim$85-120 meV), i.e., it is significantly
narrower as compared to the case of the preceding
studies.\cite{SadowskiSSC07,PlochockaPRL08} In spite of these
efforts to expand the available spectral range, a relatively weak
transmission was still found around the energy of 200~meV, due to
double-phonon absorptions in the underlying SiC substrate and
transmission spectra are affected by strong interference patterns
due to the relatively thin substrate. These two effects prevented
measurements in the energy range below the reststrahlenband of
SiC.

\section{Results and Discussion}

\begin{figure}
\scalebox{0.3}{\includegraphics*{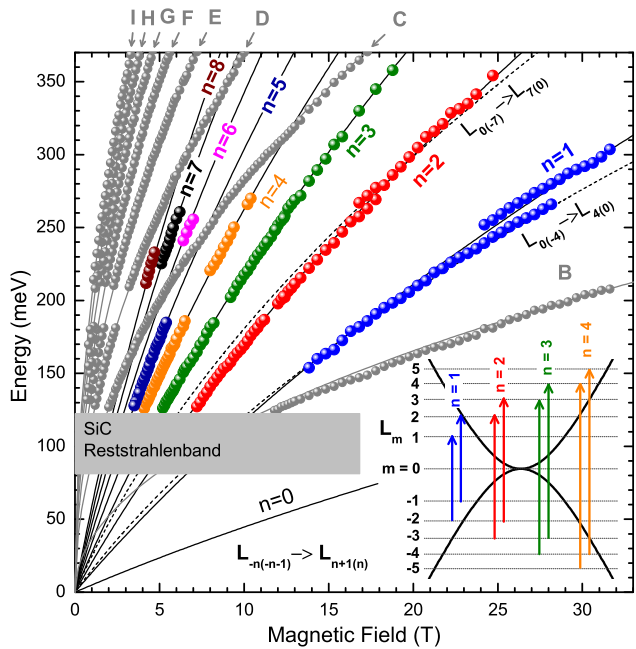}}
\caption{\label{FanChart} (color online) Fan chart: Points marked with Roman
letters, having a characteristic $\sqrt{B}$-dependence, correspond to inter-LL
transitions in electrically isolated graphene sheets.\cite{SadowskiPRL06}
Points denoted by index $n$ represent inter-band inter-LL transitions in the
graphene bilayer L$_{-n(-n-1)}\rightarrow$L$_{n+1(n)}$, as schematically shown in the inset. The full
gray lines show expected energies of transitions in the  graphene monolayer for
$v_F=1.02\times10^6$~m.s$^{-1}$, full black lines correspond to predicted positions
of the absorption lines in the graphene bilayer (only parameters
$\gamma_0=3150$~meV and $\gamma_1=385$~meV considered). The dashed lines
denote theoretical positions of two trigonal-warping-induced transitions in the
graphene bilayer L$_{0(-4)}\rightarrow$L$_{4(0)}$ and
L$_{0(-7)}\rightarrow$L$_{7(0)}$.}
\end{figure}

Typical transmission spectra of the investigated sample are shown
in Fig.~\ref{SPKT}. The dominant absorption lines which are
observed in these spectra show the characteristic
$\sqrt{B}$-dependence (see Fig.~\ref{FanChart}) and correspond to
inter-LL transitions in electrically isolated graphene sheets. We
denote those lines by Roman letters, following the initial work
and notation of Sadowski \emph{et
al.}\cite{SadowskiPRL06,SadowskiSSC07} These dominant spectral
features are equivalent to the characteristic lines observed in
the magneto-optical response of exfoliated graphene
monolayers.\cite{JiangPRL07,DeaconPRB07,HenriksenPRL10} The
subsequent absorption lines labelled here as B~$\rightarrow$~I
correspond to transitions L$_{-m(-m-1)}\rightarrow$L$_{m+1(m)}$
with $m=0\rightarrow 7$ between LLs in graphene:
$E_m=\mathrm{sign}(m)E_1\sqrt{|m|}$, where
$E_1=v_F\sqrt{2\hbar|eB|}$. The apparent Fermi velocity is
extracted to be $v_F=(1.02\pm0.02)\times10^6$~m.s$^{-1}$.
Intriguingly, the L$_{-1(0)}\rightarrow$L$_{0(1)}$ transition
exhibits a significant broadening above 16~T, which could be
tentatively related to electron-phonon interaction. This effect
will be discussed elsewhere.

The main focus of the present work are other spectral features,
i.e., the transmission dips denoted by $\textbf{n=1, n=2, \ldots
n=5}$ in Fig.~\ref{SPKT}. These absorption lines are significantly
weaker than the dominant ``graphene lines'', but are still well
resolved in our spectra. As it can be seen in Fig.~\ref{FanChart},
in contrast to the dominant transitions, these weaker lines follow
a nearly linear in $B$ dependence. As this behavior is
characteristic of massive particles and because graphene bilayer
is the simplest graphene based system with such particles, we
anticipate that electronic excitations within graphene bilayer
inclusions are responsible for the $\textbf{n=1, n=2, \ldots n=5}$
transitions. The energy ladder $\varepsilon_{n,\mu}$ of LLs in a
graphene bilayer can be easily
calculated\cite{AbergelPRB07,KoshinoPRB08} within the standard
four band model which only considers the two most relevant
coupling constants $\gamma_0$ and $\gamma_1$ (se e.g.
Ref.~\onlinecite{McCannPRL06} for their definitions):

\begin{multline}\label{Bilayer}
\varepsilon_{n,\mu}=\mathrm{sign}(n)\frac{1}{\sqrt{2}}\left[\gamma_1^2+(2|n|+1)E_1^2+\phantom{\sqrt{X}}\right.\\
\left.\mu\sqrt{\gamma_1^4 +2(2|n|+1)E_1^2\gamma_1^2+E_1^4}\right]^{1/2}.
\end{multline}

Here, a positive (negative) integer $n$ indexes the electron
(hole) LLs. $\mu=-1$ accounts for the topmost valence- and the
lowest conduction-band, whereas $\mu=1$ corresponds to two other,
split-off bands. As illustrated in the inset of
Fig.~\ref{FanChart}, optically active inter Landau level
transitions in a graphene bilayer fulfill the $|n| \rightarrow
|n|\pm 1$ selection rule. The energies of such transitions are
plotted in Fig.~\ref{FanChart} with black solid lines. Those lines
account for the transitions within the $\mu=-1$ bands. To
reproduce the experimental data, we have adjusted the $\gamma_1$
parameter whereas the Fermi velocity $v_F$ which defines $E_1$
(i.e. the intra-layer coupling $\gamma_0=3150$~meV) has been fixed
at the value derived from the monolayer-like transitions. A fair
agreement is obtained between the calculated (solid lines in
Fig.~\ref{FanChart}) and measured energies of $\textbf{n=1, n=2,
\ldots n=5}$ transitions. Optical absorptions involving LLs of
higher indexes (e.g. $\textbf{n=6, n=7}$ and $\textbf{n=8}$, see
Fig.~\ref{FanChart}) could also be observed in the spectra,
nevertheless, these lines are very weak and only visible in a
limited range of magnetic fields. Traces of inter-LL transitions
due to split-off bands ($\mu=1$ in Eq.~\ref{Bilayer}), can be also
identified in our data and experiments focused on this particular
set of transitions are in progress.

A pronounced departure of the observed bilayer transitions from
the linearity in $B$ clearly shows the limits of the parabolic
approximation which is often used for graphene bilayers in the
close vicinity of the $\mathbf{k}=0$ point, and within which the
LLs are strictly linear with the magnetic
field.\cite{McCannPRL06,NovoselovNaturePhys06} As can be seen in
Fig.~\ref{FanChart}, the positions of all these lines can be very
well reproduced with a parameter $\gamma_1=(385\pm 5)$~meV, and
these experiments thus refine the value of this parameter reported
previously from optical studies at zero magnetic
field.\cite{ZhangPRB08,KuzmenkoPRB09,ZQLiPRL09} The intriguing
splitting of the $\textbf{n=1}$ and of the $\textbf{n=2}$ lines at
high magnetic fields is beyond our simple model and will be
discussed later on.

The simplified model of LLs in the pristine bilayer graphene
provides us with reasonably accurate description of its
magneto-optical response, even though it neglects the
electron-hole asymmetry (mainly induced by tight-binding
$\gamma_4,\Delta'$
parameters),\cite{ZhangPRB08,ZQLiPRL09,KuzmenkoPRB09,ZhangNature09}
as well as a possible gap opening at the charge neutrality point.
Nevertheless, it should be noted that the optical response of the
graphene bilayer has only been unambiguously identified above the
reststrahlenband of the SiC substrate and therefore, we cannot
exclude a possible appearance of an energy gap, up to a few tens
of meV, at the $\mathbf{k}=0$ point. For the same reason, we can
only estimate a very higher limit for the carrier density in the
studied bilayer of $2\times10^{12}$~cm$^{-2}$. However, the real
carrier density is very likely similar to that of the surrounding
(electrically isolated) graphene sheets, i.e. below
$10^{10}$~cm$^{-2}$, as reported in
Refs.~\onlinecite{OrlitaPRL08II} and \onlinecite{SprinklePRL09}. We also point out
that relatively narrow linewidths of the order of 10~meV (relaxation time 
in sub-picosecond range) serve as an indication of rather high electronic quality of these  
bilayer inclusions, comparable or even better than other bilayer systems.\cite{HenriksenPRL08,OhtaScience06,RiedlPRL09}

Equivalent bilayer-like spectral features are recurrently
identified practically in all studied specimens, nevertheless,
with a strongly varying intensity. In general, we can say that the
relative intensity of these ``bilayer'' lines increases with the
total number of layers in MEG and these transitions are
practically invisible in specimens with less than 10 layers
reported in very first magneto-spectroscopy
studies.\cite{SadowskiPRL06,SadowskiSSC07} This finding serves as
an indication that we indeed observe graphene bilayer inclusions
and not regions of a local AB-stacking which might be also
speculated to appear in rotationally stacked multilayers. Such
Moir\'e-patterned AB-stacked areas have been recently visualized
in MEG by STM/STS
measurements.\cite{MillerPRB10,MillerNaturePhys10,KindermannCM10}
We further assume that twisted graphene layers which results in
the Moir\'e patterned bilayer should not provide us with so
well-defined AB stacked bilayers as we observe in our data. Let us
also note that if we compare the relative intensity of observed
transitions, we can roughly estimate that in none of the
investigated samples the ratio between bilayers and monolayers
exceeded 10\%.

We should also emphasize that the appearance of Bernal-stacked
faults in MEG, which have a form of well-defined bilayers, is not
a signature of bulk graphite. In this well known material, the
$K$-point electrons indeed mimic massive carriers in the graphene
bilayer, but with an effective inter-layer coupling $2\gamma_1$
instead of $\gamma_1$ in a real graphene
bilayer.\cite{PartoensPRB07,KoshinoPRB08,OrlitaPRL09,ChuangPRB09}
This twofold coupling in the effective bilayer model for $K$ point
electrons implies a characteristic effective mass twice enhanced
in comparison to that of massive Dirac fermions in true graphene
bilayer and consequently, also a twice lower energy separation
between adjacent interband inter-LL transitions, cf.
Fig.~\ref{FanChart} of this paper with the fan chart in
Ref.~\onlinecite{OrlitaPRL09}.

The remaining unclarified point of our study is the splitting of
the bilayer lines, which is clearly visible for transitions $\textbf{n=1}$ and $\textbf{n=2}$
around $B=17$ and 26~T, respectively. In the following,
we discuss two different scenarios for this splitting. One
possible explanation invokes the electron-hole asymmetry, reported
recently in graphene bilayers
graphene.\cite{ZhangPRB08,ZQLiPRL09,KuzmenkoPRB09} Based on this
assumption, the magnitude of the splitting for the $n$-th
transition, relative to the transition energy is expressed
by:\cite{calculation}
$$\frac{2(\Delta'/\gamma_1+2\gamma_4/\gamma_0)}{\sqrt{n(n+1)}+\sqrt{(n+1)(n+2)}
}.$$ For the values $\Delta'=0.02\:\mbox{eV}$,
$\gamma_1=0.4\:\mbox{eV}$,
$\gamma_4/\gamma_0=0.05$,\cite{ZhangPRB08,ZQLiPRL09,KuzmenkoPRB09,ZhangNature09,MakPRL09}
our measured value for $n=1$ (about 0.08) is very well reproduced.
However, the splitting due to electron-hole asymmetry should be
seen for all magnetic fields, while, as can be seen in
Fig.~\ref{SPKT}b, we only observe it in a relatively narrow range
of~$B$.

Perhaps a more natural explanation for this line splitting would
be an avoided crossing between the transition
$\mathrm{L}_{-n(-n-1)}\to\mathrm{L}_{n+1(n)}$ and some other
transition with a much smaller oscillator strength, so that it is
not seen far from the crossing point. One can see directly from
Fig.~\ref{FanChart} that the bright transitions $n=1,2$ are
crossed by the dark (i.e.,  dipole-inactive in case of a zero
trigonal warping) transitions
$\mathrm{L}_{0(-4)}\to\mathrm{L}_{4(0)}$,
$\mathrm{L}_{0(-7)}\to\mathrm{L}_{7(0)}$, respectively,
approximately at the observed values of~$B$ (the crossing occurs
at a very sharp angle, which brings a significant uncertainty).
These transitions are allowed only due to the presence of the
trigonal warping of the electronic bands, which mixes levels
$\mathrm{L}_m$ with $|m|$ differing by an integer multiple
of~$3$, see Ref.~\onlinecite{AbergelPRB07}. The ratio of the oscillator strength
of the $\mathrm{L}_{0(-4)}\to\mathrm{L}_{4(0)}$ transition to that
of the bright $n=1$ transition can be estimated~\cite{calculation}
as $(25/108)(\gamma_3\gamma_1/\gamma_0)^2(l_B/\hbar
v_F)^2\simeq0.02$ at $B=25$~T. The
$\mathrm{L}_{0(-4)}\to\mathrm{L}_{4(0)}$ transition is therefore
not expected to be seen in the experiment unless some other,
possibly resonant, admixture mechanism is taken into account.
Coupling between $\mathrm{L}_{0(-4)}\to\mathrm{L}_{4(0)}$ and
$n=1$ transitions should be quite strong as the observed
``anti-crossing splitting'' is of about 20~meV.

We have speculated this mode coupling could be due to
electron-phonon or electron-electron interactions. Electron-phonon
interaction, which could be enhanced due to the proximity of the
transition energy (250~meV) to that of the zone-center optical
phonon (196~meV), must be excluded due to the different symmetry
(this phonon is Raman active). Splitting due to Coulomb
interaction can be evaluated to be
$0.04(e^2/4\pi\varepsilon_0\hbar
v_F)(\gamma_1\gamma_3/\gamma_0)$,\cite{calculation} i.e. only
about 3~meV in the absence of dielectric screening,
$e^2/4\pi\varepsilon_0\hbar v_F=2.2$. Hence, the mechanism of the
possible strong coupling between the
$\mathrm{L}_{-1(-2)}\to\mathrm{L}_{2(1)}$ and
$\mathrm{L}_{0(-4)}\to\mathrm{L}_{4(0)}$ transitions is a puzzle
which remains to be clarified.

\section{Conclusions}

We probed graphene bilayers embedded in multilayer epitaxial
graphene on the C-terminated surface of SiC. These inclusions can
be viewed as AB-stacked faults in an otherwise
rotationally-stacked multilayer graphene structure and enable
spectroscopic studies of unperturbed graphene bilayers. The
``electronic quality'' of these bilayers is comparable or even
better than that of the bilayers obtained by exfoliation or by
epitaxial growth on the Si-terminated surface of
SiC.\cite{OhtaScience06,RiedlPRL09} This way, we could trace the
inter-band inter-LL transitions in the graphene bilayer for the
first time, and thus supply data complementary to the cyclotron
resonance absorption (i.e., intra-band inter-LL transitions)
measured on the exfoliated bilayer by Henriksen~\emph{et
al.}\cite{HenriksenPRL08} We could also clearly visualize the
departure of Landau levels in the graphene bilayer from the
linearity in $B$, which clearly sets limits for the parabolic
approximation of electronic bands in this material.

\begin{acknowledgments}
We acknowledge also funding received from EC-EuroMagNetII under Contract No. 228043,
from the Keck Foundation and from the Partner University Fund at the Embassy
of France. This work has been supported by the Projects No. 395/NPICS-
FR/2009/0, No. MSM0021620834, GACR No. P204/10/1020 and
No. MTKD-CT-2005-029671, furthermore by Grants No. 670/N-ESF-EPI/2010/0,
No. 671/NESF-EPI/2010/0, and No. GRA/10/E006 within the ESF
EuroGraphene programme (EPIGRAT).
\end{acknowledgments}


\begin{thebibliography}{42}
\bibitem[{cal()}]{calculation}
\bibinfo{note}{This expression is obtained in the two-band approximation. The
  degree of error introduced by this approximation at energies about {250 meV}
  can be roughly estimated from the deviation of the curves in Fig.~2 from
  their low-field tangents, which gives about 30\%.}
\end{thebibliography}

\end{document}